\documentclass[a4paper,fleqn,usenatbib]{mnras}

\usepackage[T1]{fontenc}
\usepackage{ae,aecompl}

\usepackage{graphicx}	
\usepackage{amsmath}	
\usepackage{amssymb}	

\title[Nitrogen]{Nitrogen evolution in the halo, thick disc, thin disc and bulge of the Galaxy}

\author[Grisoni et al.]{V. Grisoni$^{1,2}$\thanks{E-mail: valeria.grisoni@sissa.it}, F. Matteucci$^{3, 4, 2}$ \& D. Romano$^{5}$.\\
 $^1$ SISSA, Via Bonomea 265, 34136 Trieste, Italy\\  
 $^2$ INAF, Osservatorio Astronomico di Trieste, via G.B. Tiepolo 11, I-34131, Trieste, Italy\\
 $^3$ Dipartimento di Fisica, Sezione di Astronomia, Universit\`a di Trieste, via G.B. Tiepolo 11, I-34131, Trieste, Italy \\
 $^4$ INFN, Sezione di Trieste, via Valerio 2, 34134 Trieste, Italy\\
 $^5$ INAF, Osservatorio di Astrofisica e Scienza dello Spazio, Via Gobetti 93/3, 40129 Bologna, Italy\\
}

\begin{document}
\date{Accepted . ; in original form xxxx}

\pagerange{\pageref{firstpage}--\pageref{lastpage}} \pubyear{xxxx}

\maketitle

\label{firstpage}

\begin{abstract}
We study the evolution of nitrogen in the Galactic halo, thick disc, thin disc and bulge by comparing detailed chemical evolution models with recent observations. The models used in this work have already been constrained to explain the abundance patterns of $\alpha$-elements and the metallicity distribution functions of halo, disc and bulge stars; here, we adopt them to investigate the origin and evolution of N in the different Galactic components. First, we consider different sets of yields and study the importance of the various channels proposed for N production. Secondly, we apply the reference models to study the evolution of both the Galactic discs and bulge. We conclude that: i) primary N produced by rotating massive stars is required to reproduce the plateau in log(N/O) and [N/Fe] ratios at low metallicity, as well as the secondary and primary production from low- and intermediate-mass stars to reproduce the data of the thin disc. ii) The parallel model can provide a good explanation  of the evolution of N abundance in the thick and thin discs; we confirm that the thick disc has evolved much faster than the thin disc, in agreement with the results from the abundance patterns of other chemical elements. iii) Finally, we present new model predictions for N evolution in the Galactic bulge, and we show that the observations in bulge stars can be explained if massive stars rotate fast during the earliest phases of Galactic evolution, in agreement with findings from the abundance pattern of carbon.
\end{abstract}

\section{Introduction}

The understanding of the origin and evolution of nitrogen has always represented a fundamental question in Galactic Archaeology (Matteucci 1986, 2001, 2012).
\\Nitrogen has a complex nucleosynthesis and its production has both primary and secondary components, depending on whether synthesized directly from H and He or derived from metals already present in the star at birth.
Nitrogen is mostly produced by low- and intermediate-mass stars (LIMS), and it can have a secondary or primary origin. The secondary origin derives from the CNO cycle where C and O, which transform into N, are present in the star at birth. The N can be primary if the C and O are formed by the star starting from H and He. In asymptotic giant branch (AGB) stars, this primary production can occur during the phase of thermal pulses, when fresh C and O are brought to more external layers by convection and then are burned into N in the H-burning shell. This occurs during the third dredge-up followed by hot-bottom burning (see Renzini \& Voli 1981).
\\On the other hand, massive stars are producing a small fraction of the total N production and in principle it should be of secondary origin: however, Matteucci (1986) suggested that to explain the N data in Galactic halo stars, N from massive stars should have had a primary origin. In fact, such stars show a plateau of [N/Fe] at low metallicity, at variance with the secondary nature of N, which implies a continuous growth of its abundance as a function of metallicity. In the following years, Meynet \& Maeder (2002) suggested that massive stars can produce primary N if they are fastly rotating and have a very low metallicity; at very low metallicity, in fact, rotation is particularly important. The primary production of N was originally suggested by Truran \& Cameron (1971) and Talbot \& Arnett (1974), and it allows us to explain observations of N/O ratios in metal-poor Galactic dwarfs and ionized H II regions in the Milky Way and external galaxies (e.g. Sneden 1974; Smith 1975; Edmunds \& Pagel 1978; Peimbert et al. 1978; Lequeux et al. 1979; Barbuy 1983; Matteucci \& Tosi 1985 and Matteucci 1986).
\\Later on, many other theoretical studies have investigated the origin and evolution of N in the Galaxy and faced the problem of reproducing the N plateau at low metallicities (e.g. Chiappini et al. 2003, 2005, 2006, Gavil{\'a}n et al. 2006, Moll{\'a} et al. 2006, Kobayashi et al. 2011). In particular, Chiappini et al. (2005) considered the N abundances in metal-poor stars presented by Spite et al. (2005), which showed a high N/O ratio suggesting high levels of production of primary nitrogen in massive stars; in the light of those data, Chiappini et al. (2005) concluded that the only way to reproduce observations was to assume that stars at low metallicity rotate fast enough to allow massive stars to contribute large amounts of primary N. They suggested an increase in the rotational velocity in very metal-poor stars (Maeder et al. 1999, Meynet et al. 2006) and thus an increase in the yields of N (Meynet \& Maeder 2002). Moreover, Chiappini et al. (2006) tested the impact of the new stellar yields of Meynet et al. (2006) and Hirschi (2007) on the chemical evolution of CNO in the early phases of the Milky Way.
Concerning the Galactic bulge, a first attempt to follow the evolution of N was performed by Costa et al. (2005).
More recently, new sets of stellar yields were presented by Limongi \& Chieffi (2018) and implemented in Galactic chemical evolution models (Prantzos et al. 2018, Goswami et al. 2021, Romano et al. 2019). In particular, Romano et al. (2019) focused on CNO elements and concluded that their model predictions for the Milky Way can reproduce the observations if it is assumed that most stars rotate fast until a metallicity threshold is reached above which the majority of the stars have small or null rotational velocities. Recently, the question of the evolution of nitrogen has been addressed also by means of hydrodynamical cosmological simulations (Vincenzo \& Kobayashi 2018a,b).
\\From the point of view of observations, the evolution of N in the Galaxy has been studied on the basis of samples of Galactic stars, HII regions and planetary nebulae (e.g. Israelian et al. 2004, Ecuvillon et al. 2004, Rudolph et al. 2006, Costa et al. 2005; Cavichia et al. 2010, 2017, Tautvai{\v{s}}ien{\.{e}} et al. 2015, Esteban \& Garcia-Rojas 2018, Magrini et al. 2018, Horta et al. 2021, Kisku et al. 2021). For dwarf stars, observations at low [Fe/H] are challenging and there are few studies in the literature; for giant stars, the determination of N abundance is less problematic, but stellar evolution may have altered the original abundances.
Currently, we are in a golden era for this field of research thanks to the advent of large spectroscopic surveys, such as $Gaia$-ESO (Gilmore et al. 2012) and APOGEE (Majewski et al. 2017), that provide results from the bulge to the outer disc (e.g. Queiroz et al. 2020a,b). Still, there are many questions that need to be answered concerning the evolution of N (see Kobayashi et al. 2020, Roy et al. 2021, Randich \& Magrini 2021). In particular, Magrini et al. (2018) used the abundances from $Gaia$-ESO survey for Galactic field and open cluster stars, and presented observations for N in the thick and thin discs of the Galaxy, and along the disc. Furthermore, Kisku et al. (2021) presented APOGEE DR16 data for N in bulge stars. In this context, the comparison between observations and model predictions in the different Galactic components is needed to put constraints on the origin and evolution of nitrogen.
\\The aim of the present work is to model the evolution of N in the Galactic halo, thick disc, thin disc and bulge by means of detailed chemical evolution models, in the light of the most recent observational data. First, we consider the two-infall model of the Galaxy (Chiappini et al. 1997, Romano et al. 2010), to investigate N evolution over a wide metallicity range. Then, to focus on the thick and thin discs, we make use of the parallel model of Grisoni et al. (2017) (see also Chiappini 2009), that enables to study separately the evolution in the discs. This chemical evolution model has already been constrained to fit the [$\alpha$/Fe] vs. [Fe/H] plots and the metallicity distribution functions (MDFs) in the thick and thin discs (Grisoni et al. 2017, 2018), and also the abundance diagrams of various elements like lithium (Grisoni et al. 2019, 2020a, Romano et al. 2021), carbon (Romano et al. 2020), fluorine (Grisoni et al. 2020b), r- and s-process elements (Grisoni et al. 2020c), and here we adopt it to focus on the evolution of N in the thick disc and thin disc. Moreover, we will apply the model of Matteucci et al. (2019, 2020) for the Galactic bulge to follow nitrogen evolution also in this Galactic component.
\\The paper is structured as follows. In Section 2, we provide a description of the observational data used in this work. In Section 3, we outline the models adopted to follow the chemical evolution of the Galactic halo, thick disc, thin disc and bulge. In Section 4, we present the results, where we compare observations and theoretical predictions. Then, in Section 5, we outline our conclusions.

\section{Observational data}

In order to make comparisons with our theoretical predictions, in this work we consider the following observational data available for the chemical elements of interest for this work.
\\We first consider the data of Israelian et al. (2004), which include also metal-rich stars from Ecuvillon et al. (2004). They presented N abundances for both metal-poor and metal-rich dwarf stars, that span a wide range of metallicities. Their typical errors in [N/H] and [O/H] are of the order of $\sim$0.1-0.2 dex (see their Table 1 for details).
\\Then, in order to distinguish between the Galactic thick and thin disc, we take into account the recent data by Magrini et al. (2018). They presented abundances of N and O in open clusters (OCs) and field stars from $Gaia$-ESO survey (Gilmore et al. 2012, Randich et al. 2013), considering the effect of mixing in the abundance of N in giant stars and separating thick and thin disc field stars.
They have separated the field stars into thin- and thick-disc stars on the basis of their [$\alpha$/Fe] ratios, following the approach of Adibekyan et al. (2011).
Thus, we use the data by Magrini et al. (2018) to study separately the evolution in the thick disc and thin disc, and also to study the abundance gradients along the Galactic thin disc. Their typical errors on N and O abundances are $\sim$ 0.10 dex and $\sim$ 0.09 dex, respectively.
For carbon in the thick and thin discs, we consider the data of Amarsi et al. (2019). They presented C abundances from
high-resolution spectra of single dwarfs in the solar neighbourhood, after corrections for 3D and non-LTE effects. Also in this case, the separation between thin- and thick-disc stars has been done on the basis of their [$\alpha$/Fe] ratios (Adibekyan et al. 2011, 2013). For [C/Fe] abundances in thick and thin disc stars, the associated errors are of the order of $\lesssim$ 0.05 dex.
\\Finally, for N in the bulge, we take into account the recent data from Kisku et al. (2021). Their results are based on elemental abundances from Data Release 16 of the APOGEE-2 survey (Majewski et al. 2017; Ahumada et al. 2020), whose typical uncertainties are $\sim$ 0.05 dex in elemental abundances (Ahumada et al. 2020). Kisku et al. (2021) restrict their sample to have high SNR and reliable N abundances (see their Section 2.2.). To focus on the Galactic bulge, they make a spatial cut and select only stars with Galactocentric distance R $<$ 4 kpc. For more details on the observations used in this work, we refer the interested reader to the aforementioned papers.
\\In order to consistently compare the various observational data with the predictions of our chemical evolution models, we will take as reference values the solar abundances recommended by Lodders (2019).

\begin{table*}
\caption{Summary of the different yield sets used in this work. In the first column, the name of the yield set is indicated. In second column, there are the nucleosynthesis prescriptions for low- and intermediate-mass stars (LIMS) and for super asymptotic giant branch stars (super-AGB). In third column, there are the nucleosynthesis prescriptions for massive stars, and in the fourth column the corresponding rotational velocity is specified.}
\label{tab_01}
\begin{center}
\begin{tabular}{c|cccccccccc}
  \hline
\\
 Yield set & LIMS and super-AGB & Massive stars & v$_{rot}$ (km s$^{-1}$)\\
 (1) & (2) & (3) & (4) \\
\\










 \hline
000 & Ventura et al. (2013,2020) & Limongi \& Chieffi (2018) & 0 \\

 \hline

150 & Ventura et al. (2013,2020) & Limongi \& Chieffi (2018) & 150 \\

 \hline

300 & Ventura et al. (2013,2020) & Limongi \& Chieffi (2018) & 300 \\

 \hline

Var & Ventura et al. (2013,2020) & Limongi \& Chieffi (2018) & Variable$^a$ \\

 \hline

VarK & Karakas (2010) and Doherty et al. (2014a,b) & Limongi \& Chieffi (2018) & Variable$^a$ \\

 \hline

\end{tabular}

{\raggedright \textbf{Notes.} $^a$See Section 3.1.1, and Romano et al. (2019) for further details. \par}

\end{center}
\end{table*}

\section{The chemical evolution models}

In the present work, we consider the following chemical evolution models to investigate the chemical evolution of the different Galactic components:
\begin{itemize}
\item The two-infall model of the Galaxy (Chiappini et al. 1997, Romano et al. 2010);
\item The parallel model for the thick and thin discs (Grisoni et al. 2017, see also Chiappini 2009);
\item The bulge model of Matteucci et al. (2019, 2020).
\end{itemize}

\subsection{The two-infall model}

The two-infall model (Chiappini et al. 1997; Romano et al. 2010) assumes that the Milky Way formed by means of two main gas infall episodes: the first one gave rise to the halo-thick disc, whereas the second one, slower and delayed with respect to the first one, gave rise to the thin disc.
In this case, the gas infall term of a given element $i$ at the Galactocentric distance $r$ and at the time $t$ is:
\begin{align} \label{eq_2IM}
\dot G_i(r,t)_{inf}=A(r)(X_i)_{inf}e^{-\frac{t}{\tau_1}}+B(r)(X_i)_{inf}e^{-\frac{t-t_{max}}{\tau_2}},
\end{align}
where $(X_i)_{inf}$ refers to the abundance by mass for a certain element $i$ in the infalling gas. The quantities $\tau_{1}$ and $\tau_2$ are the timescales of mass accretion of the halo-thick disc and thin disc, respectively. These timescales of gas accretion are free parameters in the model, and they have been tuned in order to fit the observed metallicity distribution function in the solar vicinity (Chiappini et al. 1997, Romano et al. 2010): we consider $\tau_{1}$=1 Gyr and $\tau_2$=7 Gyr in the solar vicinity. $t_{max}$ represents the time of maximum infall on the thin disc, and it is taken equal to 1 Gyr, as in Chiappini et al. (1997) and Romano et al. (2010). The terms $A(r)$ and $B(r)$ are set to reproduce the present-time total surface mass density in the solar neighbourhood, and we consider the prescriptions of Romano et al. (2010).
\\The star formation rate (SFR) follows the Schmidt-Kennicutt law (Schmidt 1959; Kennicutt 1998a,b):
\begin{equation} \label{eq_03_02}
\psi(t) \propto \nu \sigma_{gas}^k,
\end{equation}
with $\sigma_{gas}$ being the surface gas density, k = 1.4 the index of the law, and $\nu$ the star formation efficiency (in particular, we assume $\nu =$ 2 and 1 Gyr$^{-1}$ for the halo-thick disc and thin disc, respectively).
For the initial mass function (IMF), we consider the one by Kroupa et al. (1993).

\subsection{The parallel model}

Secondly, in order to be able to follow separately the evolution in the Galactic thick and thin discs, we take advantage of the parallel model of Grisoni et al. (2017) and updated by Grisoni et al. (2019, 2020a,b,c) (see also Chiappini 2009). In this scenario, it is assumed that the thick disc and the thin disc of the Galaxy form via two distinct infall events and evolve with different rates. In this case, the gas infall term is splitted and can be writted as:
\begin{equation} \label{eq_1IMthick}
(\dot G_i(r,t)_{inf})|_{thick}=A(r)(X_i)_{inf}e^{-\frac{t}{\tau_1}},
\end{equation}
and
\begin{equation} \label{eq_1IMthin}
(\dot G_i(r,t)_{inf})|_{thin}=B(r)(X_i)_{inf}e^{-\frac{t}{\tau_2}},
\end{equation}
for the thick disc and thin disc, respectively. We adopt $\tau_1$ equal to 0.1 Gyr for the thick disc, and $\tau_{2}(r)$ 7 Gyr for the thin disc at the solar neighbourhood (Grisoni et al. 2017). The parameters $A(r)$ and $B(r)$ are set to reproduce the present-time total surface mass densities in the solar vicinity for the two discs of Nesti \& Salucci (2013). Concerning the SFR law, we assume $\nu =$ 2 and 1 Gyr$^{-1}$ for the thick disc and thin disc, respectively, and for the IMF we take the one of Kroupa et al. (1993) (see Grisoni et al. 2017).)
\\The parallel model of Grisoni et al. (2017) has been also tested at the other Galactocentric distances in Grisoni et al. (2018), where the thin disc has been divided in various concentric rings 2 kpc wide. As in Grisoni et al. (2018), we assume that the timescale for mass accretion in the Galactic thin disc varies with the Galactocentric distance according to the so-called inside-out scenario (Chiappini et al. 2001):
\begin{align} \label{eq_02a}
\tau \text{[Gyr]}=1.033 \text{r} \text{[kpc]}-1.267,
\end{align}
with the inner regions forming on a shorter timescale of formation with respect to the outer ones. Moreover, we assume a variable star formation efficiency in the Schmidt-Kennicutt law, higher in the inner regions than in the outer ones ($\nu$=8, 4, 1, 0.5 and 0.2 Gyr$^{-1}$ at a distance R of 4, 6, 8, 10 and 12 kpc from the Galactic center, respectively). On the basis of the theory of star formation induced by spiral density waves in disc galaxies (Wyse \& Silk 1989), it has been suggested the idea that the star formation efficiency varies as a function of the Galactocentric distance (see also Prantzos \& Aubert 1995, Carigi 1996, Boissier \& Prantzos 1999). In the context of chemical evolution models, it has been shown that such a variable star formation efficiency is a fundamental ingredient to properly reproduce abundance gradients (see Colavitti et al. 2009, Spitoni \& Matteucci 2011, Spitoni et al. 2015, Grisoni et al. 2018, Palla et al. 2020, Spitoni et al. 2021).

\subsection{The bulge model}

Regarding the Galactic bulge, we consider the chemical evolution model of Matteucci et al. (2019, 2020), where it is assumed that the majority of the bulge stars formed fast, on a short timescale of formation ($\tau$=0.1 Gyr) and with higher star formation efficiency ($\nu$=20 Gyr$^{-1}$) compared to the disc. Moreover, a flatter IMF than the one derived for the solar vicinity is adopted in the case of the bulge; this is consistent with expectations from the integrated galactic IMF theory (see Je{\v{r}}{\'a}bkov{\'a} et al. 2018, Yan et al. 2019 and references therein). Thus, for the Galactic bulge we use the Salpeter (1955) IMF rather than the Kroupa et al. (1993) IMF that we consider for the discs. These assumptions enable us to properly explain the [$\alpha$/Fe] ratios and MDF of bulge stars (as first suggested by Matteucci \& Brocato 1990, Ballero et al. 2007, Cescutti et al. 2009, 2018; see Barbuy et al. 2018 for a recent review on the bulge).

\begin{figure*}
\includegraphics[scale=0.53]{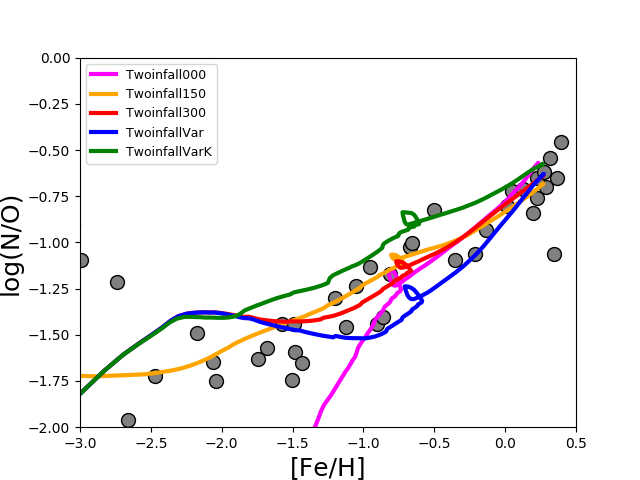}
\includegraphics[scale=0.53]{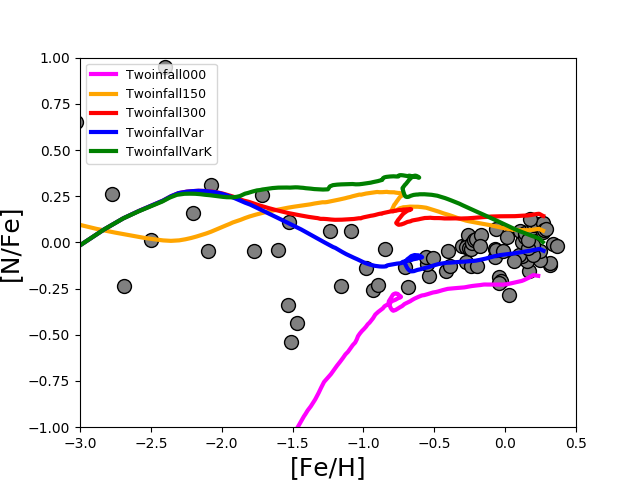}
 \caption{\textit{Left panel}: Predicted and observed log(N/O) vs. [Fe/H]. The predictions are from the two-infall model, including different yield sets as summarized in Table 1. The observational data are from the sample of Israelian et al. (2004) (grey circles), that includes also stars from Ecuvillon et al. (2004) . \textit{Right panel}: Same as the left panel, but in the case of [N/Fe] vs. [Fe/H]. In this and all the following figures, all abundance ratios are normalized to the same reference values using the solar abundances from Lodders (2019).}
 \label{fig_01}
\end{figure*}

\subsection{Nucleosynthesis prescriptions}

\begin{figure*}
\includegraphics[scale=0.53]{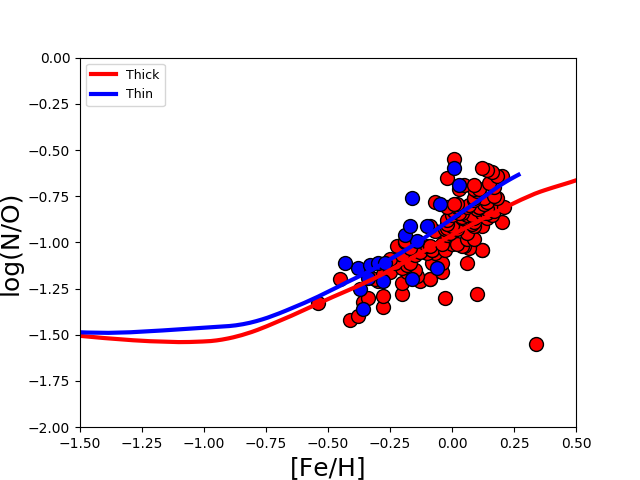}
\includegraphics[scale=0.53]{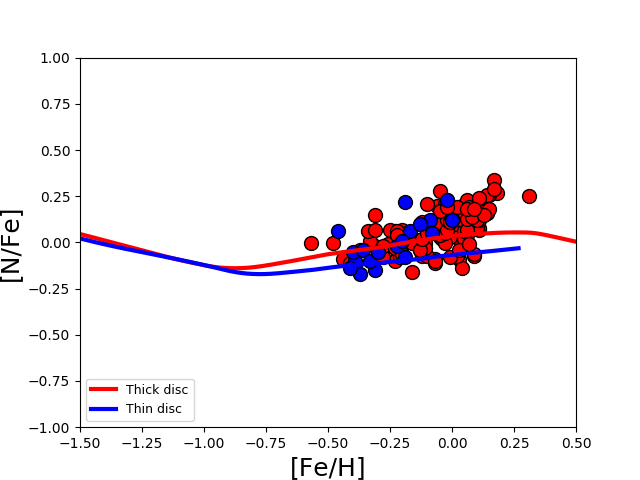}
\includegraphics[scale=0.53]{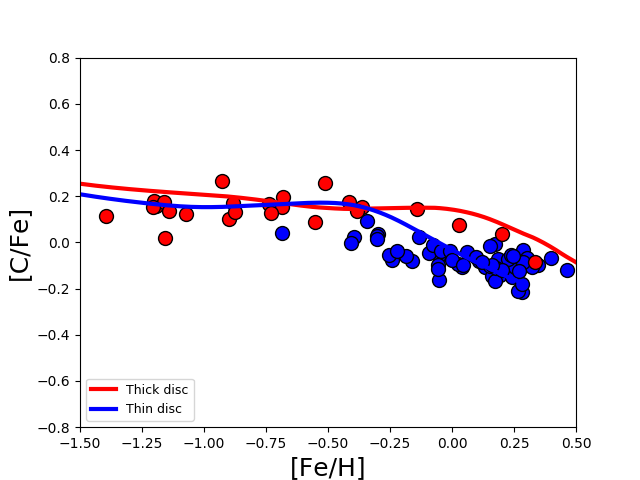}
 \caption{\textit{Upper left panel}: Predicted and observed log(N/O) vs. [Fe/H] for the thick disc and thin disc. The predictions are from the chemical evolution models of the thin disc (in blue) and thick disc (in red). The observational data are taken from Magrini et al. (2018) for the Galactic thin disc (blue circles) and thick disc (red circles). \textit{Upper right panel}: Same as the upper left panel, but in the case of [N/Fe] vs. [Fe/H]. \textit{Lower panel}: Predicted and observed [C/Fe] vs. [Fe/H] for the thick disc and thin disc. The predictions are from the chemical evolution models of the Galactic thin disc (in blue) and thick disc (in red). The observational data are taken from Amarsi et al. (2019) for the Galactic thin disc (blue circles) and thick disc (red circles). The abundance ratios have been normalized to the solar abundances recommended by Lodders (2019).}
 \label{fig_02}
\end{figure*}

The different sets of yields used in our chemical evolution models are reported in Table 1, and we provide details as follows.
In particular, we start from the best set of nucleosynthesis prescriptions for the different mass ranges by Romano et al. (2019), which tested different yield sets (see their models from MWG-01 to MWG-12) and here we consider their best ones, namely their models MWG-05,06,07,11 and 12. In particular, we update the prescriptions for low- and intermediate-mass stars and we study the effect of different rotational velocities for massive stars, since primary N is produced in rotating massive stars. In the following, we provide details about the nucleosynthesis prescriptions assumed in our chemical evolution models.
\\For low- and intermediate-mass stars (LIMS) and super-AGB stars, we adopt the nucleosynthesis prescriptions from Ventura et al. (2013) and we include also the recent yields of Ventura et al. (2020) for Z$>$0.018. The yields of Ventura et al. (2013, 2020) provide the total stellar yield and contain the dependance on metallicity; in particular, Vincenzo et al. (2016) showed how primary and secondary N for the yields of Ventura et al. vary as a function of the initial stellar mass for different metallicities and how the stellar yield of secondary N increases with metallicity (see Fig. 2 of Vincenzo et al. 2016). For LIMS and super-AGB stars, we also take advantage of the nucleosynthesis prescriptions of Karakas (2010) and Doherty et al. (2014a,b), in order to see how different prescriptions for these stars can affect the evolution of N.
\\For massive stars, we consider the yields from Limongi \& Chieffi (2018), in particular their recommended set R, with initial rotational velocities of 0, 150, and 300 km s$^{-1}$, respectively. Following Romano et al. (2019) (see their models MWG-11 and MWG-12), we adopt also the case of a variable rotational velocity for massive stars; in particular, we consider the yields of Limongi \& Chieffi (2018) computed with v$_{rot}$ = 300 km s$^{-1}$ for [Fe/H] = -3 dex and [Fe/H] = -2 dex, whereas for [Fe/H] = -1 dex and [Fe/H] = 0 dex we use the yields for non-rotating stars. In particular, we remind that primary N is produced by rotating massive stars.
\\For binary systems that give rise to SNe Ia, we adopt the stellar yields of Iwamoto et al. (1999), and for the progenitors the single-degenerate scenario is considered (see Matteucci et al. 2009).

\section{Results}

Here, we present the results, based on the comparison between observations and theoretical predictions. First of all, we consider different nucleosynthesis prescriptions in the two-infall model of the Galaxy, in order to choose the set of yields which best reproduce the evolution of nitrogen in a wide range of metallicities. Secondly, we implement the reference nucleosynthesis prescriptions in the parallel model to investigate N evolution in both the thick disc and thin disc, and along the thin disc. Finally, we present new model results for N evolution in the Galactic bulge.

\subsection{From the halo to the disc}

\begin{figure*}
\includegraphics[scale=0.53]{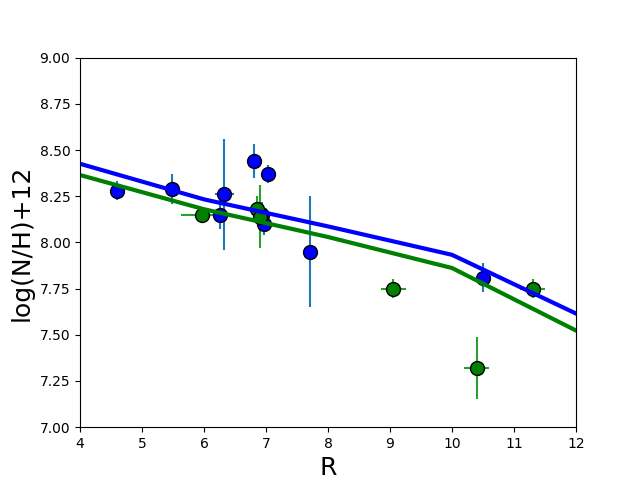}
\includegraphics[scale=0.53]{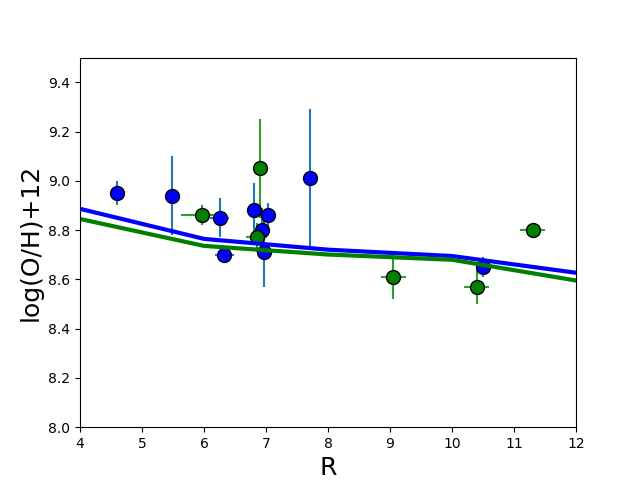}
 \caption{Predicted and observed abundance gradients along the Galactic disc. The predictions are from the reference chemical evolution model of the Galactic thin disc (the blue line represents the predictions at the present time, whereas the green line represents the prediction 2 Gyr ago). The observational data are taken from Magrini et al. (2018) and they are color-coded according to their age (the blue dots represent the young OCs, whereas the green dots represent the old ones).}
 \label{fig_03}
\end{figure*}

First, we test the different nucleosynthesis prescriptions in our chemical evolution models over a wide metallicity range and we compare with data that include a mixture of halo and disc stars. To do so, we make use of the two-infall model (Chiappini et al. 1997, Romano et al. 2010), which allows us to make predictions for the halo and the discs. This is a sequential model, with a first phase corresponding to the halo-thick disc formation and a second one corresponding to the formation of the thin disc. In particular, concerning the halo, our model aims at explaining the $in$ $situ$ component; this has been distinguished from an accreted component on the basis of chemical criteria, where high-$\alpha$ stars are supposed to form in situ in the halo, whereas low-$\alpha$ stars might have been accreted (Nissen \& Schuster 2010, Hayes et al. 2018). 
\\In the left panel of Fig. 1, we show the predicted and observed log(N/O) vs. [Fe/H]. The observational data are from Israelian et al. (2004) (grey circles), which span a wide range of metallicities and include also metal-rich stars from Ecuvillon et al. (2004). The predictions are from the two-infall model of the Galaxy, including different yield sets as indicated in Table 1.
In all the models, we have a first phase corresponding to the halo-thick disc formation, then we have a gap due to the assumed threshold in the star formation process, which marks the transition with the second phase corresponding to the thin disc formation: this gap is visible as a small loop in the model predictions of Fig. 1. In the halo phase, rotating massive stars can set a plateau in both log(N/O) and [N/Fe] at low metallicities (both the case with v$_{rot}$=150 and 300 km s$^{-1}$) at variance with the case where zero rotational velocity is assumed, and this is because rotating massive stars produce primary N. On the other hand, the second phase corresponding to the thin disc formation is mostly dominated by the contribution of low- and intermediate-mass stars: we show both the results obtained by including the yields of Ventura et al. (2013,2020) and the ones of Karakas (2010) and Doherty et al. (2014a,b).
From the comparison between model predictions and observational data over a wide metallicity range, we can conclude that the model with the yields computed by including a variable rotational velocity from massive stars of Chieffi \& Limongi (2018) and the yields of Ventura et al. (2013,2020) for LIMS provides the best agreement with observations, and we will take these nucleosynthesis prescriptions as reference in the following calculations.
\\In summary, rotation produces primary N and a variable rotational velocity is required in order to better reproduce the observational data in the whole metallicity range, in agreement with the findings of Romano et al. (2019) for CNO isotopes, Romano et al. (2020) for C, and also Grisoni et al. (2020b) for fluorine. The reference model can provide also a solar value (log(N/H)+12 equal to 7.96) in agreement with the obervations recommended by Lodders (2019) (i.e. 7.85 $\pm$ 0.12).

\subsection{Thick and thin discs}

Then, in order to focus on the Galactic thick and thin discs and to follow separately the evolution in these two components, we make use of the parallel model of Grisoni et al. (2017) and we compare with data that distinguish between thick-disc and thin-disc stars. In the thick and thin disc models, we apply the reference nucleosynthesis prescriptions that have been previously established over a wide metallicity range, namely the yield set with the yields of Ventura et al. (2013,2020) for LIMS and super-AGB and the ones of Limongi \& Chieffi (2018) with variable rotational velocity (see Table 1).
\\In Fig. 2, we show the predicted and observed log(N/O) vs. [Fe/H] plot (upper left panel) and [N/Fe] vs. [Fe/H] plot (upper right panel) for the thick disc and thin disc of the Galaxy. The predictions are from the chemical evolution models of the Galactic thin disc (in blue) and thick disc (in red). The observational data are taken from Magrini et al. (2018) for the Galactic thin disc (blue circles) and thick disc (red circles).
We can distinguish two different sequences corresponding to the Galactic thick disc and thin disc, both from the observations and from the model predictions, even if they are less evident than in the abundance patterns of other elements, such as Mg (Grisoni et al. 2017). This is because N and Fe have more similar timescales of production than Mg and Fe.
In fact, $\alpha$-elements such as Mg are mainly produced by core-collapse SNe on a short timescale, at variance with Fe which is mainly produced by SNe Ia on a longer timescale, and this fact produces the typical feature in the [$\alpha$/Fe] vs. [Fe/H] that can be interpreted in terms of the so-called time-delay model (Matteucci 2001, 2012). In the [$\alpha$/Fe] vs. [Fe/H] plot, the dichotomy between the thick disc and thin disc is evident, with the thick disc being $\alpha$-enhanced compared to the thin disc, due to a shorter timescale of formation and higher star formation efficiency.
Thus, we assume that the thick disc formed with a shorter timescale of formation and a higher star formation efficiency, as suggested by studies based on the abundance patterns of the $\alpha$-elements (Grisoni et al. 2017, 2018), and we compare our model predictions with the observations in the case of N.
In the [N/Fe] vs. [Fe/H] plot, the thick-disc model predicts a decrease at the high metallicities, as we will see later for the bulge. We remind that the thick-disc model predicts very few stars forming at those metallicities since the thick disc formed fast reaching minimal star formation very rapidly (see Grisoni et al. 2017); still, it is interesting to see the model predictions at high metallicities, even if further data would be required to draw firm conclusions about this feature.
\\Since N and C are strictly related, in the lower panel of Fig. 2, we present also our predictions for the [C/Fe] vs. [Fe/H] plot compared to the recent data by Amarsi et al. (2019). From the lower panel of Fig. 2, it can be seen that the bimodality of the thick disc and thin disc is more evident both from the data and from the models, as it happens for Mg (Grisoni et al. 2017), with the thick disc having higher [C/Fe] values compared to the thin disc at a certain metallicity. Concerning carbon, there have been also other observational studies that presented the [C/Fe] vs. [Fe/H] plot in the thick disc and thin disc; first, Bensby \& Feltzing (2006) studied C in the thick and thin discs, but did not find a clear dichotomy between the two components. Then, Griffith et al. (2019) and Franchini et al. (2020) presented observations of C in the thick and thin discs from GALAH and Gaia-ESO survey, respectively. In Romano et al. (2020), these various observational trends for C and the variation of C abundance in galaxies have been studied in detail, and we address the interested reader to the Romano et al. (2020) paper.
\\In conclusion, the parallel model can provide a good explanation  of the evolution of N abundance in the thick and thin discs; we conclude that the thick disc had a much faster evolution with respect to the thin disc, in accordance with results based on the abundance patterns of other chemical elements (Grisoni et al. 2017, 2019, 2020a,b,c; Romano et al. 2020).

\subsection{Abundance gradients}

After the study of N evolution in the solar vicinity, we investigate abundance gradients along the Galactic thin disc.
\\In Fig. 3, we show the abundance gradient of N along the Galactic thin disc.
The predictions are from the reference chemical evolution models of the Galactic thin disc, where we implement the prescriptions of the reference model of Grisoni et al. (2018) for the various Galactocentric distances, where the abundance gradients and their evolution with time have been studied in detail. Thus, we assume inside-out and variable star formation efficiency, which are fundamental ingredients to reproduce abundance gradients (see also Palla et al. 2020, Spitoni et al. 2021).
The observational data are from Magrini et al. (2018) and they are color-coded according to their age: the blue dots represent the young OCs (ages $<$ 1 Gyr), whereas the green dots represent the old ones (ages $\ge$ 1 Gyr). We note that to improve the agreement between data and model predictions, it would require to fine-tune the inside-out law or the variable SFE law proposed by Grisoni et al. (2018), in order to best fit the data of Magrini et al. (2018), but we can see that the overall behaviour can be reproduced by those assumptions. Similarly, in the right panel of Fig. 3, we show also the abundance gradient of oxygen along the Galactic thin disc, which is also available from the sample of Magrini et al. (2018). We note that the gradient of O, both from the theory and from the data, is flatter than the one of N, due to the different production channels of the two elements: in fact, O is mainly produced by core-collapse SNe on short timescales, at variance with N where an important contribution comes from long-lived stars.
\\In conclusion, our model can reproduce not only the abundance patterns in the solar vicinity, but also the abundance gradients along the Galactic disc by assuming the inside-out scenario and a variable star formation efficiency, as suggested in Grisoni et al. (2018).

\begin{figure*}
\includegraphics[scale=0.53]{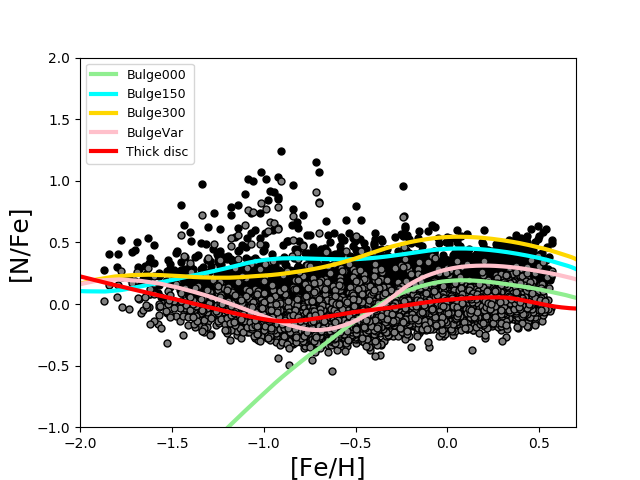}
 \caption{Predicted and observed [N/Fe] vs. [Fe/H] for the Galactic bulge. The predictions are from the chemical evolution models of the Galactic bulge, including different yield sets as summarized in Table 1, and they are compared to the reference model for the thick disc. The observational data are taken from Kisku et al. (2021), both uncorrected (black dots) and corrected (gray dots) for the mixing effect by a factor of -0.25 dex, as suggested by Magrini et al. (2018). The abundance ratios have been normalized to the solar abundances recommended by Lodders (2019).}
 \label{fig_04}
\end{figure*}

\subsection{Bulge}

Finally, we study N evolution in the Galactic bulge, in the light of the most recent observational data.
\\In Fig. 4, we show the observed and predicted [N/Fe] vs. [Fe/H] for the Galactic bulge. The observational data for bulge stars are from Kisku et al. (2021): we show both the data uncorrected (black dots) and corrected (gray dots) for the mixing effect by a factor of -0.25 dex, as suggested by Magrini et al. (2018). In fact, we checked that APOGEE data should be in this range to be consistent with the Gaia-ESO ones presented by Magrini et al. (2018) and corrected for the mixing effect. The predictions are from the chemical evolution model of the Galactic bulge of Matteucci et al. (2019, 2020), which assumes a shorter timescale of formation, higher star formation efficiency and a flatter IMF with respect to the disc. For comparison, we also show the curve corresponding to our reference model of the thick disc. The different curves for the bulge correspond to different rotational velocities for the yields of Limongi \& Chieffi (2018): variable rotational velocity, and v$_{rot}$ equal to 0, 150 and 300 km s$^{-1}$. The results reinforce the idea that that massive stars should rotate fast during the early phases of the evolution of the Galaxy. In this context, Romano et al. (2020) investigated the evolution of C in the Galactic bulge and they showed that, for fast-rotating massive stars (with a rotational velocity of 300 and 150 km s$^{-1}$), they match very well the [C/Fe] and [C/O] bulge data; on the contrary, the adoption of non-rotating stellar yields leads to severely underestimate the [C/Fe] ratio. Thus, our conclusions for the Galactic bulge are in accordance with the results of Romano et al. (2020) on the basis of (very few) observations for carbon. A scenario where the first stellar generations have been fast rotators has been suggested also by Cescutti et al. (2018) by means of their stochastic chemical evolution model for the bulge.
\\In this context, we note that a fundamental role is played by the IMF, which should favor the formation of massive stars in regimes of bursting star formation (see Romano et al. 2017, Zhang et al. 2018). In particular, for the bulge model we assume a Salpeter (1955) IMF rather than the one of Kroupa et al. (1993) that we assume for the disc. Thus, in the Galactic bulge, N production from massive stars is enhanced by the adoption of an IMF flatter than the canonical one in the high-mass domain: in this way, for the bulge, we can get higher [N/Fe] ratios with respect to the disc. In general, top-heavy IMFs are requested to reproduce other observational properties of spheroids (Matteucci 2012) and they are consistent with the IGIMF theory (Je{\v{r}}{\'a}bkov{\'a} et al. 2018; Yan et al. 2019, and references therein).
\\Thus, in summary, we reinforce the idea that massive stars rotate fast during the earliest phases of Galactic evolution and we emphasize the important role played by the IMF in this context.

\section{Conclusions}

In this work, we studied the evolution of N in the Galactic halo, thick disc, thin disc and bulge by using detailed chemical evolution models and comparing our theoretical predictions with recent observations. We considered differents sets of nucleosynthesis prescriptions for the different mass ranges by Romano et al. (2019) (their models MWG-05,06,07, 11 and 12), and we included also the new yields of Ventura et al. (2020) for metal-rich AGB stars. Then, i) we adopted the two-infall model to study N evolution over a wide metallicity range; ii) we applied the parallel model by Grisoni et al. (2017) to focus on the evolution of N in the Galactic thick disc and thin disc, iii) we studied the abundance gradients along the Galactic thin disc by applying the prescriptions of Grisoni et al. (2018), and finally iv) we applied the model of Matteucci et al. (2019) for the Galactic bulge to investigate the evolution of N also in this Galactic component in the light of recent observations.
\\The conclusions of this paper are summarized in the following way.
\begin{itemize}
\item We analyse the effect of rapidly rotating massive stars by adopting the yields of Limongi \& Chieffi (2018) in the two-infall model of the Galaxy. We show that they can reproduce the plateau at low metallicities, since they produce only primary N. Previous works adopting stellar rotation only at very low metallicity have faced the problem of reproducing the data for N (e.g. Chiappini et al. 2006). We show that the best agreement with the observational data of N in the whole metallicity range is obtained by considering a variable rotational velocity, where most massive stars rotate fast during the early phases of the evolution of the Galaxy, while they rotate much less at later times. These conclusions are in agreement with the recent findings for carbon (Romano et al. 2020) and fluorine (Grisoni et al. 2020b).
\item Then, in order to focus on the Galactic thick and thin discs and to study separately these two components, we take advantage of the parallel model of Grisoni et al. (2017), where we implement the reference nucleosynthesis prescriptions as previously established. We conclude that the thick disc had a much faster evolution with respect to the thin disc, in accordance with results for other chemical elements, like the $\alpha$-elements (Grisoni et al. 2017, Spitoni et al. 2019, 2021), lithium (Grisoni et al. 2019, 2020a, Romano et al. 2021), carbon (Romano et al. 2020), fluorine (Grisoni et al. 2020b) and neutron-capture elements (Grisoni et al. 2020c).
\item In the [N/Fe] vs. [Fe/H] plot, the bimodality between the Galactic thick disc and thin disc is less evident than in the [Mg/Fe] vs. [Fe/H] one (Grisoni et al. 2017), since N and Fe have more similar timescales of production than Mg and Fe.
\item The previous assumptions have allowed us to reproduce also the observed gradients along the Galactic thin disc, by assuming the inside-out scenario for disc formation and a variable star formation efficiency along the disc (in agreement with Grisoni et al. 2018).
\item Finally, concerning the Galactic bulge, we consider the chemical evolution model of Matteucci et al. (2019), which assumes a shorter timescale of accretion, a more efficient star formation and a flatter IMF with respect to the discs. We conclude that the observations in bulge stars can be explained if massive stars rotate fast during the early phases of the evolution of the Galaxy, in agreement with results from the abundance pattern of carbon (Romano et al. 2020).
In this context, a fundamental role is played by the IMF, which should favor the formation of massive stars in regimes of bursting star formation (see Romano et al. 2017, Zhang et al. 2018).

\end{itemize}

In conclusions, N is mostly produced by LIMS as secondary and partly as primary, but rotating massive stars are important in the earliest phases of Galactic evolution and they can explain the N plateau at low metallicities as due to the production of primary N as a consequence of rotation which induces a mechanism for which fresh C and O are brought into the H-burning shell where N is manufactured.

\section*{Acknowledgments}

We thank the anonymous referee for useful comments and suggestions that improved the paper.
\\We acknowledge Shobhit Kisku and Ricardo Schiavon for kindly providing the data for nitrogen in the bulge.
\\V.G. acknowledges financial support at SISSA from the European Social Fund operational Programme 2014/2020 of the autonomous region Friuli Venezia Giulia. D.R. is grateful to the International Space Science Institute (ISSI, Bern, CH, and ISSI-BJ, Beijing, CN) for funding the team "Chemical abundances in the ISM: the litmus test of stellar IMF variations in galaxies across cosmic time" (PIs D. Romano, Z.-Y. Zhang).

\section*{Data availability}

The derived data generated in this research will be shared on reasonable request to the corresponding author.

\end{document}